\documentstyle[12pt,epsfig]{article}
\begin{document}
\baselineskip=23pt \vspace{1.2cm}
\begin{center}{\large \bf EFFECT  OF ISOVECTOR-SCALAR MESON ON
 NEUTRON \\STAR MATTER  IN STRONG MAGNETIC FIELDS }
\bigskip
\\
F.X.~Wei\footnote{E-mail: weifx@ihep.ac.cn}$^{1}$,
G.J.~Mao$^{2,3}$, C.~M. Ko$^{4}$,
L.S.~Kisslinger$^{5}$,\\
H.~St\"{o}cker$^{6}$, and W.~Greiner$^{6}$
\bigskip
\\
{\em $^1$Institute of High Energy Physics,Chinese Academy of Sciences,\\
 P.O.Box 918(4), Beijing 100049, China\footnote{mailing address}\\
$^2$CCAST (World Laboratory), P.O.Box 8730, Beijing 100080,
 China\\
 $^3$Department of Physics, Beihang University, Beijing 100083, China\\
 $^4$Cyclotron Institute and Physics Department, Texas A-M
 University, \\ College Station, Texas 77843-3366\\
 $^5$Department of Physics, Carnegie Mellon University,
 Pittsburgh  PA  15213\\
 $^6$Frankfurt Institute for Advanced Studies, Johann Wolfgang
 Goethe-University,\\
 Max-von-Laue-Str. 1, D-60438 Frankfurt am Main, Germany}
\end{center}
\bigskip
\begin{abstract}
\par
We study the effects of isovector-scalar meson $\delta$ on the
equation of state (EOS) of neutron star matter in strong magnetic
fields. The EOS of neutron-star matter and nucleon effective
masses are calculated in the framework of Lagrangian field theory,
which is solved within the mean-field approximation. From the
numerical results one can find that the $\delta$-field leads to a
remarkable splitting of proton and neutron effective masses. The
strength of $\delta$-field decreases with the increasing of the
magnetic field and is little at ultrastrong  field. The proton
effective mass is highly influenced by magnetic fields, while the
effect of magnetic fields on the neutron effective mass is
negligible. The EOS turns out to be stiffer at $B < 10^{15}$G but
becomes softer at stronger magnetic field after including the
$\delta$-field. The AMM terms can affect the system merely at
ultrastrong magnetic field($B
> 10^{19}$G). In the range of $10^{15}$ G -- $10^{18}$ G
 the properties of neutron-star matter are found to be
similar with those without magnetic fields.
\end{abstract}
PACS numbers: 26.60.+c, 21.65.+f \\

\section{Introduction}
\hspace*{\parindent} In the standard relativistic mean field (RMF)
model \cite{wal74,ser86} of nuclear matter, the $\sigma$ ,
$\omega$ and $\rho$ meson  are used in descriptions of nuclear
interactions. The short range of the isovector-scalar meson
$a_{0}$(980)(the $\delta$ meson) exchange justifies neglecting its
contribution to symmetric nuclear matter. However, for strongly
isospin-asymmetric matter at high densities in neutron stars, the
contribution of the $\delta$-field should be considered. Resent
theoretical studies\cite{sch96,liu02,kub97,men04,li04} motivate
the investigation of the effect of the isovector-scalar meson on
the neutron-star matter. They have found \cite{liu02} that the
$\delta$-field leads to a large repulsion in dense neutron-rich
matter and definite splitting of proton and neutron effective
masses. The energy per particle of neutron matter becomes larger
at high densities than the one with no $\delta$-field included and
the proton fraction of $\beta$-stable matter
increases\cite{kub97}. Those properties play an important role in
description of the structure and stability conditions of neutron
stars. A splitting of proton and neutron masses can affect the
transport properties of dense matter \cite{li05}. As is well
known, there are strong magnetic fields of $10^{14}$G \cite{mel99}
on the neutron star surface. The strength of magnetic fields in
the interior of neutron stars can be up to $10^{18} G$
 \cite{ler77}. The neutron-star matter in strong magnetic fields
without the isovector-scalar field has been studied, which give
interesting and novelty results \cite{bro00,car01,cha97}. It is
our main aim to investigate the $\delta$-field influence on the
properties of the neutron-star matter in the presence of magnetic
fields.
\par
Theoretical studies about the effects of very strong magnetic
fields on the EOS of neutron-star matter indicated\cite{bro00}
that the softening of the EOS caused by Landau quantization was
overwhelmed by stiffening due to the incorporation of the
anomalous magnetic moments of nucleons. At high baryon densities,
muons can be produced in the charge-neutral , beta-equilibrated
matter with respect to the channel of $ e^{-} \longleftrightarrow
\mu^{-} + \nu_{e} + \overline{\nu}_{\mu}$ , as soon as the
chemical potential of electrons $\mu_{e}$ reaches a value equaling
to the muon rest mass. In cold neutron stars neutrinos and photons
already escape and the chemical potentials of those can be set as
zero. Consequently, we get $\mu_{e} = \mu_{\mu}
 $($\mu_{\mu}$ is the chemical potential of muons). As the baryon
density increases, the densities of muons and electrons  become
comparable with that of nucleons. The inclusion of the anomalous
magnetic moments of leptons thus makes sense. In this paper we
will study the effects of the AMM of nucleons and leptons in a
dense neutron-star matter including the $\delta$-field effect.
\par
In the following section, the Lagrangian field theory of
interacting nucleons and mesons including magnetic fields will be
introduced. The numerical results are given in section 3. We
separate the cases with and without the inclusion of the AMM
effects. The modification of proton and neutron effective masses
in the dense matter with strong magnetic fields will be discussed
in detail. The energy per particle and EOS will be studied too. In
Section 4, we summarize our results and prospect for the possible
descriptions of additional components such as hyperons and quarks.
\section{Formalism}
\hspace*{\parindent}The application of Lagrangian field theory to
the study of neutron stars was first carried out by
Glendenning\cite{gle.1}. We consider a neutron-star matter
consisting of neutrons, protons, electrons and muons in the
presence of a uniform magnetic field B along the $z$-axis. The
Lagrangian density can be written as
\begin{eqnarray}
{\cal L}& = & \bar{\psi}_{b}[i\gamma_{\mu}\partial^{\mu} -
q_{b}\frac{1 + \tau_{0}}{2}\gamma_{\mu}A^{\mu} -
\frac{1}{4}\kappa_{b}\mu_{N}\sigma_{\mu\nu}F^{\mu\nu} - M_{b} +
g_{\sigma}\sigma +
g_{\delta}\mbox{\boldmath{$\tau$}}\cdot\mbox{\boldmath{$\delta$}}
- g_{\omega}\gamma_{\mu}\omega^{\mu} -
g_{\rho}\gamma_{\mu}\mbox{\boldmath $\tau$}\cdot{\bf R}^{\mu}]\psi_{b} \nonumber\\
 &&+\bar{\psi_{l}}[i\gamma_{\mu}\partial^{\mu} - q_{l}\gamma_{\mu}A^{\mu} -
\frac{1}{4}\kappa_{l}\mu_{B}\sigma_{\mu\nu}F^{\mu\nu} -
m_{l}]\psi_{l} +
\frac{1}{2}\partial_{\mu}\sigma\partial^{\mu}\sigma - U(\sigma) -
\frac{1}{2}m_{\sigma}^{2}\sigma^{2} +
\frac{1}{2}\partial_{\mu}\mbox{\boldmath$\delta$}\partial^{\mu}\mbox{\boldmath$\delta$}
\nonumber \\
&&- \frac{1}{2}m_{\delta}^{2}\mbox{\boldmath$\delta$}^{2} -
\frac{1}{4}\omega_{\mu\nu}\omega^{\mu\nu} +
\frac{1}{2}m_{\omega}^{2}\omega_{\mu}\omega^{\mu} -
\frac{1}{4}{\bf R}_{\mu\nu}\cdot{\bf R}^{\mu\nu} +
\frac{1}{2}m_{\rho}^{2}{\bf R}_{\mu}{\bf R}^{\mu} -
\frac{1}{4}F_{\mu\nu}F^{\mu\nu},
\end{eqnarray}
where $\psi_{b}$ and $\psi_{l}$ are the baryon(b = n, p) and
lepton(l = e, $\mu$) fields; $\sigma$, $\omega_{\mu}$, {\bf R},
\mbox{\boldmath$\delta$} represent the scalar meson, vector meson,
isovector-vector meson and isovector-scalar meson field, which are
exchanged for the description of nuclear interactions. $A^{\mu}
\equiv (0$, $0$, $Bx$, $0)$ refers to a constant external magnetic
field along the $z$-axis. The field tensors for the $\omega$,
$\rho$ and magnetic field are given by $\omega_{\mu\nu} =
\partial_{\mu}\omega_{\nu} - \partial_{\nu}\omega_{\mu}$, ${\bf R}_{\mu\nu}
= \partial_{\mu}{\bf R}_{\nu} - \partial_{\nu}{\bf R}_{\mu}$ and
$F_{\mu\nu} = \partial_{\mu}A_{\nu} - \partial_{\nu}A_{\mu}$ .
 U($\sigma$) is the self-interaction part of the scalar
field\cite{bog77}: $U(\sigma) = \frac{1}{3}b \sigma^{3} +
\frac{1}{4}c \sigma^{4}$.
\par
$M_{b}$ and $m_{l}$ are free baryon masses and lepton masses, and
$m_{\sigma}$, $m_{\omega}$, $m_{\rho}$, $m_{\delta}$ are the
masses of the $\sigma$, $\omega$, $\rho$ and $\delta$ meson,
respectively. $\mu_{N}$ and $\mu_{B}$ are the nuclear magneton of
nucleons and Bohr magneton of electrons; $\kappa_{p}$ = 3.5856,
$\kappa_{n}$ = -3.8263, $\kappa_{l}$ = $\frac{\alpha_{l}}{\pi}$
and $\alpha_{l}$ = $1159652188(4) \times 10^{12} (4ppb)$,
$\alpha_{\mu}$ = $11659203(8) \times
10^{-10}(0.7ppm)$\cite{mao03,ben02,byr03} are the coefficients of
AMMs for protons, neutrons, electrons and leptons, respectively.
Then the anomalous magnetic moment can be defined by the coupling
of the baryons and leptons to the electromagnetic field tensor
with $\sigma_{\mu\nu}$ = $\frac{i}{2} [\gamma_{\mu}$,
$\gamma_{\nu}]$ and $\kappa_{i}$($i = n, p, e,\mu$) given above.
\par
The field equations in a mean field approximation(MFA), in which
the meson fields are replaced by their expectation values in a
many-body ground state, are given by
\begin {eqnarray}
&&{(i\gamma_{\mu}\partial^{\mu} - q_{b}\frac{1 +
\tau_{0}}{2}\gamma_{\mu}A^{\mu} -
\frac{1}{4}\kappa_{b}\mu_{N}\sigma_{\mu\nu}F^{\mu\nu} - m_{b}^{*}
- g_{\omega}\gamma_{\mu}\omega^{\mu} -
g_{\rho}\gamma_{0}\tau_{3b}R_{0}^{0})}\psi_{b} = 0 ,\\
&&{(i\gamma_{\mu}\partial^{\mu} - q_{l}\gamma_{\mu}A^{\mu} -
\frac{1}{4}\kappa_{b}\mu_{N}\sigma_{\mu\nu}F^{\mu\nu} - m_{l}
)}\psi_{l} = 0 ,\\
&&{\ \ \ \ }m_{\sigma}^{2}\sigma + b\sigma^{2} + c\sigma^{3} = g_{\sigma}\rho_{s},\\
&&{\ \ \ \ }m_{\omega}^{2}\omega_{0} = g_{\omega}\rho_{b},\\
&&{\ ~ ~ ~ }m_{\rho}^{2}R_{00} = g_{\rho}(\rho_{p} - \rho_{n}),\\
&&{\ ~ ~ ~ }m_{\delta}^{2}\delta_{0} = g_{\delta}(\rho_{s}^{p} -
\rho_{s}^{n}).
\end{eqnarray}
The energy-momentum tensor can be written as\\
\begin{eqnarray}
T_{\mu\nu} & =
&{i\bar{\psi}_{b,l}\gamma_{\mu}\partial_{\nu}\psi_{b,l}} +
g_{\mu\nu}[\ \frac{1}{2}m_{\sigma}^{2}\sigma^{2} + U(\sigma) +
\frac{1}{2}m_{\delta}^{2}\ \mbox{\boldmath$\delta$}^{2} -
\frac{1}{2}m_{\omega}^{2}\omega_{\lambda}\omega^{\lambda}\nonumber \\
&&- \frac{1}{2}m_{\rho}^{2}{\bf R}_{\lambda}{\bf R}^{\lambda} +
\frac{B^{2}}{2}] + \partial_{\nu}A^{\lambda}F_{\lambda\mu}.
\end{eqnarray}
Here  $\rho_{b} = \rho_{p} + \rho_{n}$ is the baryon number
density, and  $\rho_{s} = \rho_{s}^{p} + \rho_{s}^{n}$ is the
scalar number density. In the mean field approximation, the first
two components of the isospin vector ${\bf R}_{\mu}$  and
\mbox{\boldmath$\delta$} vanish; i.e, $\langle{{\bf
R}^{\mu}}\rangle = \langle{R^{\mu}_{0}}\rangle$;
$\langle{\mbox{\boldmath$\delta$}}\rangle =
\langle{\delta_{0}}\rangle$.
The effective baryon masses are thus expressed as\\
\begin{eqnarray}
m_{p}^{*} &=& M_{p} - g_{\sigma}\sigma - g_{\delta}\delta_{0} ,\\
m_{n}^{*} &=& M_{n} - g_{\sigma}\sigma + g_{\delta}\delta_{0}.
\end{eqnarray}
In the presence of the AMM of nucleons and leptons, the energy
spectra of particles can be expressed as \\
\begin{eqnarray}
E_{\nu,s}^{p} &=& \sqrt{k_{z}^{2} + (\sqrt{2eB\nu + m_{p}^{*2}} +
s\Delta_{p})^{2}} + g_{\omega}\omega_{0} + g_{\rho}R_{0,0} , \\
E_{s}^{n} &=& \sqrt{ k_{z}^{2} + (\sqrt{k_{x}^{2} + k_{y}^{2} +
m_{n}^{*2}} + s\Delta_{n})^{2}} + g_{\omega}\omega_{0} - g_{\rho}R_{0,0}, \\
E_{\nu,s}^{l} &=& \sqrt{k_{z}^{2} +  (\sqrt{2eB\nu + m_{l}^{2}} +
s\Delta_{l})^{2}},
\end{eqnarray}
where  $\Delta_{b} = -\frac{1}{2}\kappa_{b}\mu_{N}B$ and
$\Delta_{l} = -\frac{1}{2}\kappa_{l}\mu_{B}B$.
\par The number densities of protons, neutrons and leptons read as \\
\begin{eqnarray}
\rho_{p} &=&
\frac{eB}{2\pi^{2}}[\sum_{\nu=0}^{\nu_{max}}k_{f,\nu,1}^{(p)}
+ \sum_{\nu=1}^{\nu_{max}}k_{f,\nu,-1}^{(p)}],\\
\rho_{n} &=& \frac{1}{2\pi}\sum_{s}\{\frac{2}{3}k_{f,s}^{(n)3} +
s\Delta_{n}[(m_{n}^{*} + s\Delta_{n})k_{f,s}^{(n)} +
E_{f}^{(n)2}(\arcsin\frac{m_{n}^{*} + s\Delta_{n}}{E_{f}^{(n)}} -
\frac{\pi}{2})]\},
\end{eqnarray}
and\\
\begin{equation}
\rho_{l} =
\frac{eB}{2\pi^{2}}[\sum_{\nu=0}^{\nu_{max}}k_{f,\nu,1}^{(l)} +
\sum_{\nu=1}^{\nu_{max}}k_{f,\nu,-1}^{(l)}],
\end{equation}
respectively. In the above expressions, $k_{f,\nu,s}^{(p)}$ and
$k_{f,\nu,s}^{(l)}$ are the Fermi momenta of protons and leptons
for the landau level $\nu$ and the spin index $s = -1, 1$;
$k_{f,s}^{(n)}$ is the Fermi momentum of neutrons.  They are
related to the Fermi energies as
\begin{eqnarray}
k_{f,\nu,s}^{(p)}  &=&  \sqrt{E_{f}^{(p)2} - (\sqrt{ m_{p}^{*2} +
2eB\nu} + s\Delta_{p})^{2}} , \\
k_{f,s}^{(n)}  &=&  \sqrt{E_{f}^{(n)} - (m_{n}^{*} +
s\Delta_{n})^{2}}
\end{eqnarray}
and\\
\begin{equation}
k_{f,\nu,s}^{(l)}  =  \sqrt{E_{f}^{(l)2} - (\sqrt{ m_{l}^{2} +
2eB\nu} + s\Delta_{l})^{2}}.
\end{equation}
The scalar number densities of nucleons have the form of\\
\begin{eqnarray}
\rho_{s}^{p} & = &
\frac{eBm_{p}^{*}}{2\pi^{2}}\left[\sum_{\nu=0}^{\nu_{max}}\frac{\sqrt{m_{p}^{*2}
+ 2eB\nu} + \Delta_{p}}{\sqrt{m_{p}^{*2} + 2eB\nu}}{\rm ln}\left|
\frac{k_{f,n,1}^{(p)} +
E_{f}^{(p)}}{\sqrt{m_{p}^{*2} + 2eB\nu} + \Delta_{p}}\right|\right. \nonumber\\
& + &  \left.\sum_{\nu=1}^{\nu_{max}}\frac{\sqrt{m_{p}^{*2} +
2eB\nu} - \Delta_{p}}{\sqrt{m_{p}^{*2} + 2eB\nu}}{\rm ln}\left|
\frac{k_{f,n,-1}^{(p)} +
E_{f}^{(p)}}{\sqrt{m_{p}^{*2} + 2eB\nu} - \Delta_{p}}\right|~\right],\\
 \rho_{s}^{n} & = &
\frac{m_{n}^{*}}{4\pi^{2}}\sum_{s}\left[k_{f,s}^{(n)}E_{f}^{(n)} -
(m_{n}^{*} + s\Delta_{n})^{2}{\rm ln}\left|\frac{k_{f}^{(n)} +
E_{f}^{(n)}}{m_{n}^{*} + s\Delta_{n}}\right|~\right].
\end{eqnarray}
The energy densities of nucleons and leptons are
given as\\
\begin{eqnarray}
\varepsilon_{p} & = & \frac{1}{4\pi^{2}}
\sum_{\nu=0}^{\nu_{max}}\left[k_{f,\nu,1}^{(p)}E_{f}^{(p)} +
(\sqrt{m_{p}^{*2} + 2eB\nu} + \Delta_{p})^{2}{\rm ln}\left|
\frac{k_{f,\nu,1}^{(p)} +
E_{f}^{(p)}}{\sqrt{m_{p}^{*2} + 2eB\nu} + \Delta_{p}}\right|~\right]\nonumber \\
&& +
\frac{1}{4\pi^{2}}\sum_{\nu=1}^{\nu_{max}}\left[k_{f,\nu,-1}^{(p)}E_{f}^{(p)}
+ (\sqrt{m_{p}^{*2} + 2eB\nu} - \Delta_{p})^{2}{\rm ln}\left|
\frac{k_{f,\nu,-1}^{(p)} +
E_{f}^{(p)}}{\sqrt{m_{p}^{*2} + 2eB\nu} - \Delta_{p}}\right|~\right],\\
\varepsilon_{n}  & = &
\frac{1}{4\pi^{2}}\sum_{s}\left\{\frac{1}{2}k_{f,s}^{(n)}E_{f}^{(n)3}
+ \frac{2}{3}s\Delta_{n} E_{f}^{(n)3}(\arcsin\frac{m_{n}^{*} +
s\Delta_{n}}{E_{f}^{(n)}} - \frac{\pi}{2})  +
(\frac{s\Delta_{n}}{3} -
\frac{m_{n}^{*} + s\Delta_{n}}{4})\right.\nonumber \\
&&\times\left.\left[(m_{n}^{*} +
s\Delta_{n})k_{f,s}^{(n)}E_{f}^{(n)} + (m_{n}^{*} +
s\Delta_{n})^{3}{\rm ln}\left|\frac{k_{f}^{(n)} +
E_{f}^{(n)}}{m_{n}^{*} +
s\Delta_{n}}\right|~\right]\right\}, \\
\varepsilon_{l} & = & \frac{1}{4\pi^{2}}
\sum_{\nu=0}^{\nu_{max}}\left[k_{f,\nu,1}^{(l)}E_{f}^{(l)} +
(\sqrt{m_{l}^{2} + 2eB\nu} + \Delta_{l})^{2}{\rm
ln}\left|\frac{k_{f,\nu,1}^{(l)} +
E_{f}^{(l)}}{\sqrt{m_{l}^{2} + 2eB\nu} + \Delta_{l}}\right|~\right]\nonumber\\
&& +
\frac{1}{4\pi^{2}}\sum_{\nu=1}^{\nu_{max}}\left[k_{f,\nu,-1}^{(l)}E_{f}^{(l)}
+ (\sqrt{m_{l}^{2} + 2eB\nu} - \Delta_{l})^{2}{\rm ln}\left|
\frac{k_{f,\nu,-1}^{(l)} +
E_{f}^{(l)}}{\sqrt{m_{l}^{2} + 2eB\nu} - \Delta_{l}}\right|~\right].\\
\end{eqnarray}
\par
The total energy density of the n-p-e-$\mu$ system is\cite{bro00} \\
\begin{equation}
  \varepsilon = \varepsilon_{p} +  \varepsilon_{n} +
\varepsilon_{e} + \varepsilon_{\mu} +
 \frac{1}{2}m_{\sigma}^{2}\sigma^{2} + U(\sigma) +
\frac{1}{2}m_{\delta}^{2}\delta_{0}^{2} +
\frac{1}{2}m_{\omega}^{2}\omega_{0}^{2} + \frac{1}{2}m_{\rho}
^{2}R_{0,0}^{2} + \frac{B^{2}}{8\pi}.
\end{equation}
where the last term is the contribution from the external magnetic
field.
\par
 Because of the charge neutrality and chemical
equilibrium the pressure of the system can be obtained by
\begin{equation}
P = \sum_{i}\mu_{i}\rho_{i} - \varepsilon = \mu_{n}\rho_{b} -
\varepsilon.
\end{equation}

\section{Numerical results}
\hspace*{\parindent}From above expressions, one can obtain the
nucleon effective masses and the EOS of the system after solving
the meson field equation, of (4)--(7), numerically. The chemical
equilibrium conditions of $\mu_{n}$ = $\mu_{p}$ + $\mu_{e}$ and
$\mu_{e}$ = $\mu_{\mu}$, as well as the charge neutrality
$\rho_{p}$ = $\rho_{e}$ + $\rho_{\mu}$ are applied in the
iteration procedure. The nucleon-meson coupling constants and the
coefficients in the scalar field self-interactions are obtained by
adjusting to the bulk properties of symmetric nuclear matter where
the magnetic field is absence. Furthermore, for symmetric nuclear
matter the leptons are omitted and protons and neutrons have equal
densities. The saturation properties consist
 of the binding energy$(E/A)$, compression modulus$(K)$, symmetry energy$(a_{sym})$,
 the effective mass$(m_{N}^{*}/M_{N})$ and the pressure $p$. The binding energy can be obtained
by $E/A = \varepsilon/\rho_{b} - M_{b}$. The symmetry energy
reads\cite{kub97}
\par
\begin{equation}
 a_{sym} = \frac{1}{2}C_{\rho}^{2}\rho_{0} +
\frac{k_{f}^{2}}{6\sqrt{k_{f}^{2} + m^{*2}}} -
C_{\delta}^{2}\frac{m^{*2}\rho_{0}}{2(k_{f}^{2} + m^{*2})(1
+ C_{\delta}^{2}A(k_{f},m^{*}))},\\
\end{equation}
where \\
\begin{equation}
A(k_{f}, m^{*}) =
\frac{4}{(2\pi)^{3}}\int_{0}^{k_{f}}\frac{k^{2}d^{3}k }{(k^{2} +
m^{*2})^{3/2}}
\end{equation}
is a function of the Fermi momentum, $k_{f} = k_{f}^{(p)} =
k_{f}^{(n)}$, and the effective mass, $m^{*} = m_{p}^{*} =
m_{n}^{*}$. For symmetric nuclear matter at saturation density
$\rho_{0}$ , we have defined  $C_{\sigma} =
g_{\sigma}/m_{\sigma}$, $C_{\omega} = g_{\omega}/m_{\omega}$,
$C_{\rho} = g_{\rho}/m_{\rho}$, $C_{\delta} =
g_{\delta}/m_{\delta}$. In the presence of $\delta$-fields, the
compression modulus is calculated to be
\begin{eqnarray*}
\frac{1}{9}K &=& k_{f}^{2}\frac{\partial^{2}}{\partial
 k_{f}^{2}}(\frac{\varepsilon_{0}}{\rho})\mid_{\rho
= \rho_{0}}\nonumber \\
&=& C_{\omega}^{2}\rho_{0} +
\sum_{N}\frac{k_{f}^{2}}{6E_{f}^{(N)}} + \frac{\rho_{0}}{2} -
\frac{(B_{p} + B_{n})^{2} + C_{\delta}^{2}[f_{\sigma}(B_{p} -
B_{n})^{2} + 2A_{n}B_{p}^{2} + 2A_{p}B_{n}^{2}]}{2f_{\sigma} +
C_{\delta}^{2}[(A_{n} + A_{p})f_{\sigma} + 2A_{n}A_{p}] + A_{n} +
A_{p}},
\end{eqnarray*}
where N = n, p, and
\begin{eqnarray*}
A_{N} &=& \frac{6\rho_{s}^{N}}{m_{N}^{*}} -
\frac{3\rho_{0}}{E_{N}},\\
B_{N} &=& \frac{m_{N}^{*}}{E_{N}}, {\ \ \ } f_{\sigma} =
\frac{U(\sigma)}{g_{\sigma}^{2}}.
\end{eqnarray*}
\par
Recently, improved empirical data for the compressibility and
symmetry energy are available \cite{vre03,col04,sou04}, which
benefit to the study of evident isospin asymmetric nuclear matter.
In this work we adopt the parameter sets obtained in
Ref.\cite{liu02}. The coupling constants and the corresponding
saturation properties of nuclear matter are listed in Table 1. One
can see that SetA with and without the $\delta$-field produce the
same saturation properties, which fits well into the range of new
empirical data. Therefore, it is suitable to be used to
investigate the $\delta$-field effect in isospin asymmetric
matter. Another set of parameters GM3\cite{gle.2} is widely used
in neutron-star matter calculations. The results for neutron-star
matter without magnetic fields are presented in Fig.1.
\par
The dotted lines in Fig. 1(a) denote the effective masses of
neutrons and protons as functions of baryon densities in a
n-p-e-$\mu$ system reckoned in the model of SetA including the
$\delta$-field. The splitting of proton and neutron masses can
affect the transport properties of dense matter\cite{kub97}. The
difference between the proton and the neutron effective mass
inclines to decrease with the increasing of the baryon density. It
indicates that the strength of $\delta$-field $- g_{\delta}\delta$
decreases with densities. For comparison the results for the model
of SetA without $\delta$-field and set GM3 are presented in the
figure too. In Fig. 1(b) and Fig. 1(c) we show the pressure
densities and energy densities as functions of the baryon density.
The deviation between the results of SetA($NL\sigma\omega\rho$)
and SetA($NL\sigma\omega\rho\delta$) are negligible. Although the
proton fraction increases by including the
$\delta$-field\cite{liu02}, the EOS has little change due to the
fact that the energy density and pressure density display the
average contributions of protons and neutrons, and the effects of
isospin vector fields are small and cancel somewhat.
Alternatively, because of a larger effective mass the results of
Set GM3 deviate from SetA evidently. In the following
calculations, we present the numerical results under strong
magnetic fields by separating the cases with and without the
inclusion of the AMM effects.
\subsection{Effect of the magnetic fields without AMM terms }
\hspace*{\parindent}We consider magnetic fields in the range of
$10^{12}$- $10^{20}$G. There are two characteristic strengths of
magnetic fields for the problem involved, which are critical
fields for electrons($B_{c}^{e} = 4.414 \times 10^{13}$G) and
protons($B_{c}^{p} = 1.487 \times 10^{20}$G)\cite{bro00}. We are
more interested in the results at the vicinity of $B = B_{c}^{e}$
and ultrastrong magnetic fields around $B = 10^{18}$G. Therefore,
in the following calculations we consider magnetic fields of $B =
10^{12}$G , $10^{13}$G, $10^{15}$G, $10^{5} \times B_{c}^{e}$,
$10^{19}$G, $3 \times 10^{19}$G. For comparison, the results for
$B = 0$ will be presented too. In order to manifest the influence
of magnetic fields on neutron-star matter, in the results of
pressure and energy density given below the magnetic energy part
will be excluded.
\par
Figure 2 depicts the nucleon effective masses as functions of
baryon densities for various magnetic fields. From Fig. 2(a) it
can be seen that the effective mass of protons at $B = 10^{12}G$,
$10^{13}$G is much larger than that without magnetic field. The
results of $B = 10^{15}$G, $10^{5}\times B_{c}^{e}$, $10^{19}$ G
are indistinguishable from those of $B = 0$G. When the field
strength further increases, a smaller $m_{p}^{*}$ was obtained.
The enhancement and suppression of proton effective masses mainly
results from the variation of proton fraction caused by the effect
of magnetic fields. The effective masses of neutrons given in Fig.
2(b), however, have little changes. The different response of the
proton and neutron effective mass can be explained in Figure 3,
where the meson field strength as functions of baryon densities
are displayed. Fig. 3(a) gives the $\sigma$-field strength
-$g_{\sigma}\sigma$ as a function of the density. One can find
that the curves behave similar to that of the proton effective
mass, except that the $m_{p}^{*}$ varies more rapidly due to the
including of the $\delta$-field. Fig. 3(b) delineates that the
$\delta$-field strength changes substantially with magnetic fields
around $B = 10^{13}$G. The nucleon effective masses are entirety
defined by the $\sigma$- and $\delta$-field. The proton effective
mass is enhanced significantly at $B = 10^{12}$G and $B =
10^{13}$G because of the large cancellation between the $\sigma$-
and $\delta$-field. With the increasing of the magnetic field, the
strength of the $\sigma$- and $\delta$-field approaches the field
free case, so does the proton effective mass. The situation for
neutron effective masses is different. The including of the
$\delta$-field lowers the effect of magnetic fields on the
$\sigma$-field. Thus the neutron effective mass is almost
independent with magnetic fields. One can also see that at larger
magnetic field($B > 10^{15}$G) and higher density the
$\delta$-field strength is near zero. That indicates that the
effective masses of neutrons and protons tend to be same at higher
density. At ultrastrong magnetic field($B = 3 \times 10^{19}$G),
the $\delta$-field strength turns out to be negative at density of
$\rho > 2\rho_{0}$. We also present the strength of
$\omega$-field($g_{\omega}\omega_{0}$) and
$\rho$-field($g_{\rho}R_{0,0}$) in Fig. 3(c) and Fig. 3(d),
respectively. The $\omega$-field strength is solely defined by the
baryon density and thus linear with it. The $\rho$-field
represents the isospin asymmetry of nuclear matter. From Fig.
3(d), one can find that the $\rho$-field strength for magnetic
fields of $B = 10^{12}$G and $10^{13}G$ are much larger than that
for others, which means that the proton fractions are very small
and the neutron-star matter are extremely asymmetric. At
ultrastrong magnetic fields $(B = 3 \times 10^{19})$ the
$\rho$-field is almost zero, i.e, the density of protons is
approximately equal to that of neutrons. The neutron-star matter
inclines to isospin symmetric at very large magnetic field.
\par
Figure 4 shows the EOS of neutron-star matter under magnetic
fields for SetA(NL$\sigma\omega\rho\delta$) with the AMM terms
excluded. For pressure and energy density we present the matter
part only. In Fig. 4(a), the energy per nucleon as a function of
the baryon density is depicted for various magnetic fields. ,
while the pressure is given in Fig. 4(b) and (c). One can see that
the equation of state becomes stiffer around B = $10^{12}$G
compared to the field-free case. At the range of $B = 10^{15}$G --
$10^{19}$G the EOS is indistinguishable to that of $B = 0$. At
even larger magnetic field the EOS comes out to be softer which is
in accordance with previous studies\cite{bro00}. The variation of
EOS with magnetic field strength can be understood by the particle
fraction as discussed before. The results of
SetA$(NL\sigma\omega\rho)$ without $\delta$-field are shown in
Figure 5. The general trends of the nucleon effective mass, the
$\rho$-field and EOS are similar to that as depicted in Fig. 2 and
Fig. 4, but the magnitude of variation is suppressed. Thus one may
conclude that the magnetic field effects change the proton
fraction and the $\delta$-field enhances the isospin asymmetric
effects.
\par
From above, we can find that the $\delta$-field leads to splitting
of nucleon effective masses. Under magnetic field the proton
effective mass decrease rapidly with magnetic field and in range
of $10^{15}G < B < 10^{19}$G the results is almost
indistinguishable from that of $B=0$. The neutron effective mass
have little change for different fields because the effects of
magnetic field on $\sigma$-field and $\delta$-field counteract.
The EOS change a lot for different magnetic field because of the
change of proton fraction. By including of $\delta$-field, the EOS
of neutron-star matter is stiffer for strong magnetic fields ($B <
10^{15}$G) and become softer for ultrastrong magnetic fields ($B >
10^{18}$G). The effect of $\delta$-field decrease with magnetic
field and is little at magnetic field of $B > 10^{18}$G. The EOS
for $10^{15} < B < 10^{19}$G is similar with that of $B = 0$ and
at ultrastrong magnetic field($B > 10^{19}$G) the neutron-star
matter tends to symmetric. From above analysis, we also find that
the fraction of proton play an important role in the description
of nucleon effective mass and the EOS of neutron-star matter.
\subsection{ Effect of AMM terms}
\hspace*{\parindent}In previous works\cite{bro00}, the effect of
AMM terms of nucleons on the EOS of $n-p-e-\mu$ system in the
absence of $\delta$-field are studied in detail. In our studies,
we will introduce the AMM term of muons. At the magnetic field $B
= 10^{17}$-$10^{20}$G, the densities of muons are comparable with
baryon densities\cite{bro00}. The effect of the muon AMM term is
then expected to make sense. We adopt the value $a_{\mu} =
1165203(8)\times 10^{-10}(0.7ppm)$\cite{ben02,byr03}, which is a
present word average experimental value. In the following we show
the results including the AMM terms of nucleons and muons, while
the effect of the electron AMM term will be investigated in the
next section.
\par
The proton effective mass as a function of the baryon density is
shown in Fig. 6(a). One can find that the results for $B <
10^{19}$G have very little change compared with the case without
AMM terms. For $B = 3\times 10^{19}$G the proton effective mass no
more decreases monotonically but tends to reach certain situation
at high density. A similar situation is exhibited for the neutron
effective mass as shown in Fig. 6(b). AT ultra high magnetic field
the $m_{n}^{*}$ becomes larger than the field-free case while for
lower B they are indistinguishable. This is mainly caused by the
effects of AMM terms on the $\sigma$-field. In Figure 7 one can
see that the scalar field is increased at large field, especially
for high densities. The changes on the $\delta$- and $\rho$-field
are negligible for with and without AMM terms.
\par
The EOS of the system are shown in Figure 9. Again, the main
modifications due to the including of AMM terms come out for high
fields of $B > 10^{19}$G. The energy per nucleon becomes strongly
binding at lower densities. Alternately, a stiffer pressure is
displayed compared to the case without AMM terms. It should be
mentionable that  the EOS is only for matter, the magnetic energy
itself has not yet been added.
\par
We can conclude from above that at ultrastrong magnetic fields($B
> 10^{19}$G) the proton effective masse and neutron effective
mass all become larger by including of AMM terms. And the
effective masse of neutron is bigger than that of proton effective
masses at lower densities ($\rho< 4 \rho_{0}$), since in this
regions the density of protons is larger than that of neutrons.
The EOS becomes much stiffer at $B = 10^{19}$G because of the
effect of AMM terms. The nucleon effective mass and EOS of this
system have no evident change under magnetic fields of $B <
10^{15}$G. The AMM terms play an important role only under very
strong magnetic field of $B > 10^{18}$G. and the effect of AMM
terms decrease with magnetic field. On the other hand, the effect
of $\delta$-field decreases with magnetic field and is very small
at $B > 10^{18}$G. Its main contribution was shown at $B \sim
10^{12}G$
\par
\subsection{ Results including AMM of electrons}
\hspace*{\parindent}The critical field of electrons is about
$10^{13}$G. Most of magnetic field strengths considered in this
work are around or well beyond this point. It is generally
believed that the high-order terms stemming from the vacuum
polarization of electrons in an external magnetic field
\cite{dun00} may get into work near the critical point and cancel
the electron AMM term. Nevertheless, it is interesting to check
the effects of the electron AMM term numerically in the present
system. The results including the $\delta$-field and AMM terms of
all relevant particles are depicted in Figure 9. Fig. 9(a) and (b)
show the nucleon effective masses as functions of baryon densities
. The pressure as functions of baryon densities and energy
densities are show in Fig. 10(c). One can see that the nucleon
effective masses and EOS have negligible changes compared with the
situation excluding the AMM of electrons.
\section{Summary and outlook}
\hspace*{\parindent}We have studied the properties of the
neutron-star matter consisting of n-p-e-$\mu$ in strong magnetic
fields. For nuclear interactions we applied the relativistic mean
field theory with the exchange of $\sigma$-, $\omega$-, $\rho$-
and $\delta$-mesons. Our main interest is to investigate the
influences of isospin vector field on the asymmetric matter in the
presence of magnetic fields. The effects of AMM terms of nucleons
and leptons are included. Two sets of coupling constants with and
without $\delta$-field are used in calculations. The nucleon
effective masses and EOS are studied in detail. Interesting
results have been found for two regions of magnetic field
strength. At lower field of $B \sim 10^{12}$G, where the AMM terms
play no role, the proton effective mass was enhanced significantly
compared to the case of $B = 0$. The equation of state becomes
much stiffer. This is mainly caused by the change of proton
fraction. The neutron effective mass is almost independent of
magnetic fields because the effect of $\sigma$- and $\delta$-field
cancel to some extent. In the range of $B 10^{15}--10^{18}$G, no
obvious differences were found both on the nucleon effective
masses and EOS for with and without the magnetic field. At larger
field of $B \sim 10^{19}$G, the proton effective mass increases
with magnetic field at high density($\rho > 5\rho_{0}$) while at
lower density it becomes less than the neutron effective mass. The
EOS of neutron-star matter is softer at ultrastrong field but
becomes stiffer with the inclusion of AMM terms. Besides, one can
find that the neutron-star matter tends to be symmetric at the
range calculated. Compared with the results without
$\delta$-field, we find that the effect of $\delta$-field
decreases with magnetic field and becomes little at $B >
10^{19}$G. It can also be found that the effect of AMM terms
increases with magnetic field and is very little when the field
strength $B < 10^{15}$G. At the end, we have presented the results
including $\delta$-field and AMM of nucleons, muons and electrons,
and find that the effect of the electron AMM is very little.
Particularly, the proton fraction can be proved to play an
important role in descriptions of properties of neutron-star
matter .
\par
The vector self-interaction terms of $\omega$-field can influence
the maximum mass, the rotational frequency and cooling properties
of neutron stars\cite{gle.3}. The spin polarization of protons
probably affect the structure and composition of neutron stars.
These questions will be studied in forthcoming work. With the
densities increasing, the hyperon and quark degrees of freedom
must be considered for the core of neutron stars
 \cite{sch96,gle.3}.  The interactions of quarks are very deferent
 from that of nucleons and can lead to many new results\cite{wang05}.
 All these can influence
the EOS of neutron star matter and warrant further studies.
\par

\textbf{Acknowledgements:} The authors are grateful to N. Van.
Giai, J. Schaffner and B.~Liu for fruitful discussions. This work
was supported by the National Natural Science Foundation of China
under Grant No. 10275072.

\newpage
\par
\par
\renewcommand{\baselinestretch}{1.6}
{\tiny
\begin{tabular}{cccccccccccc}
\multicolumn{12}{c}{TABLE 1}\\
\multicolumn{12}{c}{\scriptsize Parameter sets and the
corresponding
saturation properties of nuclear matter}\\
\hline \hline Parameter Sets&
$C_{\sigma}^{2}$&$C_{\omega}^{2}$&$C_{\rho}^{2}$&$C_{\delta}^{2}$&b&c
 &$\rho_{0}$& E/A&$m_{N}^{*}/M_{N}$ & $a_{sym}$ & K \\
 & (fm) & (fm) & (fm) &(fm)&$(fm^{-1})$&&$fm^{-3}$ &(MeV) && (MeV)&(MeV) \\
setA(NL$\rho$)&10.32924&5.42341&0.94999&0.0000&0.03302&-0.00483&0.16&-16.0&0.75&31.3&240\\
SetA(NL$\rho\delta$)&10.32924&5.42341&3.1500&2.5000&0.03302&-0.00483&0.16&-16.0&0.75&31.3&240\\
GM3&9.927&4.820&1.198&0.000&0.041205&-0.002421&0.153&-16.3&0.78&32.5&240\\\hline
\end{tabular}}
\vspace{\bigskipamount}

\newpage
\begin{figure}[h]
\label {fig1} \centering
\mbox{\epsfig{figure=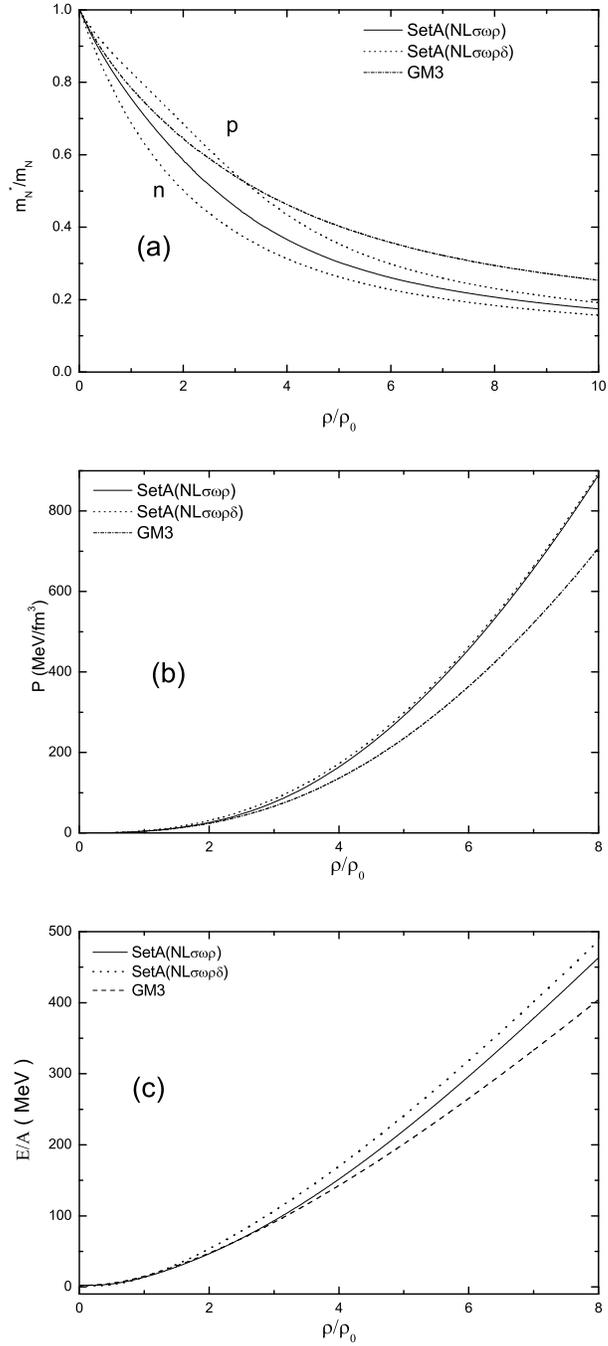,height=0.9\textheight, clip=}}
\caption{Nucleon effective masses $m_{N}^{*}/M_{N}$(a) and
pressure(b) as functions of baryon densities $\rho/\rho_{0}$ for
the neutron-star matter without magnetic fields; (c) shows the
energy per nucleon as a function of baryon density.}
\end{figure}

\newpage
\begin{figure}[h]
\label {fig2} \centering
\mbox{\epsfig{figure=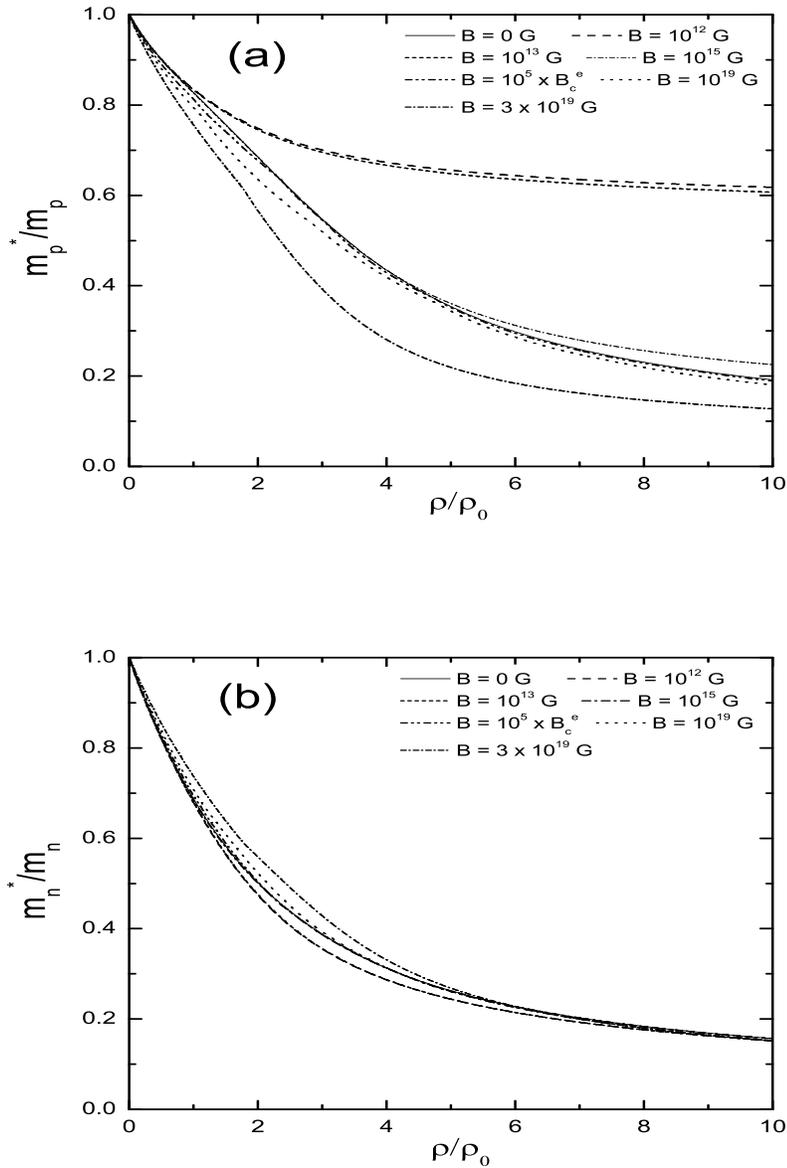,height=0.8\textheight, clip=}}
\caption{ Effective masses of protons(a) and neutrons(b) as
functions of the density for different magnetic field strengths.
The parameter SetA($NL\sigma\omega\rho\delta$) has been used in
calculations. }
\end{figure}

\newpage
\begin{figure}[h]
\label {fig3} \centering
\mbox{\epsfig{figure=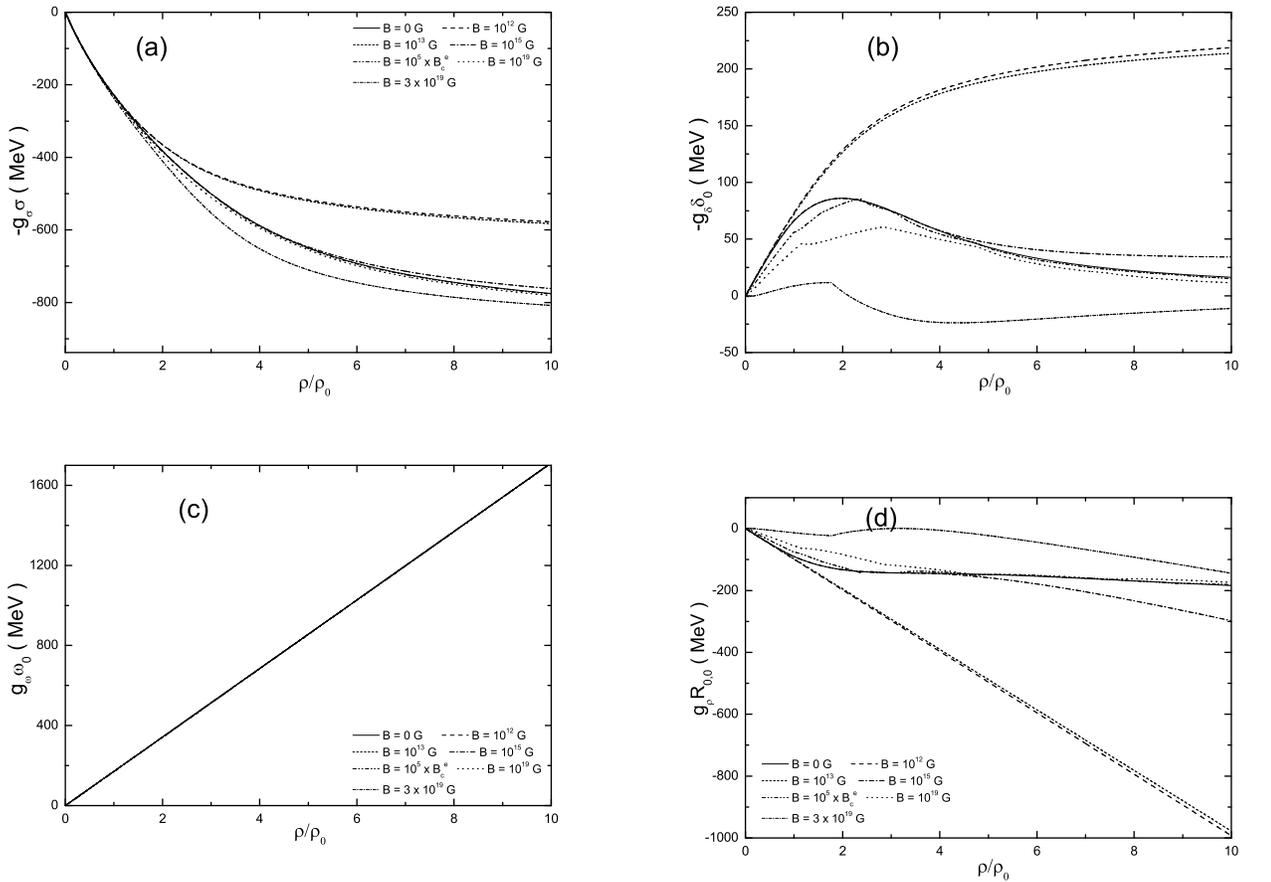,height=0.65\textheight, clip=}}
\caption{The strength of the $\sigma$-field(a), $\delta$-field(b),
$\omega$-field(c) and $\rho$-field(d) as functions of densities
for different magnetic fields.}
\end{figure}

\newpage
\begin{figure}[h]
\label {fig4} \centering
\mbox{\epsfig{figure=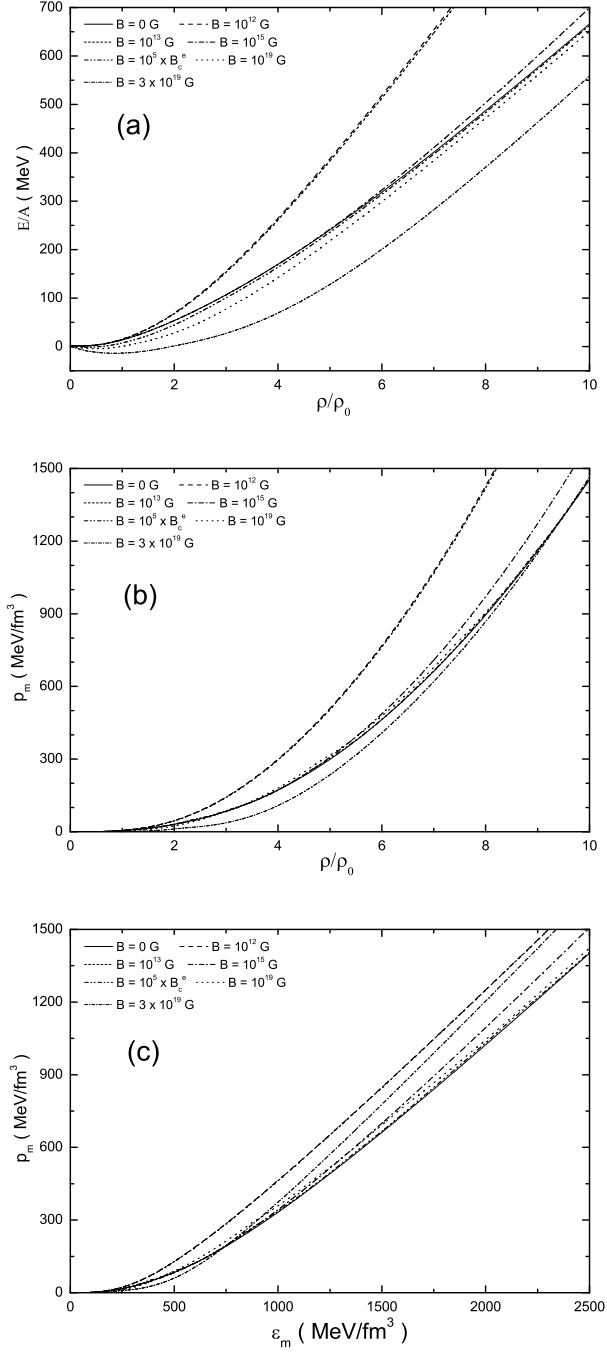,height=0.9\textheight, clip=}}
\caption{ Energy per nucleon(a) and the matter part of pressure
density $p_{m}$(b) as functions of the baryon density, Figure(c)
shows the $p_{m}$ as a function of matter energy densities
$\varepsilon_{m}$.}
\end{figure}

\newpage

\begin{figure}[h]
\label {fig5} \centering
\mbox{\epsfig{figure=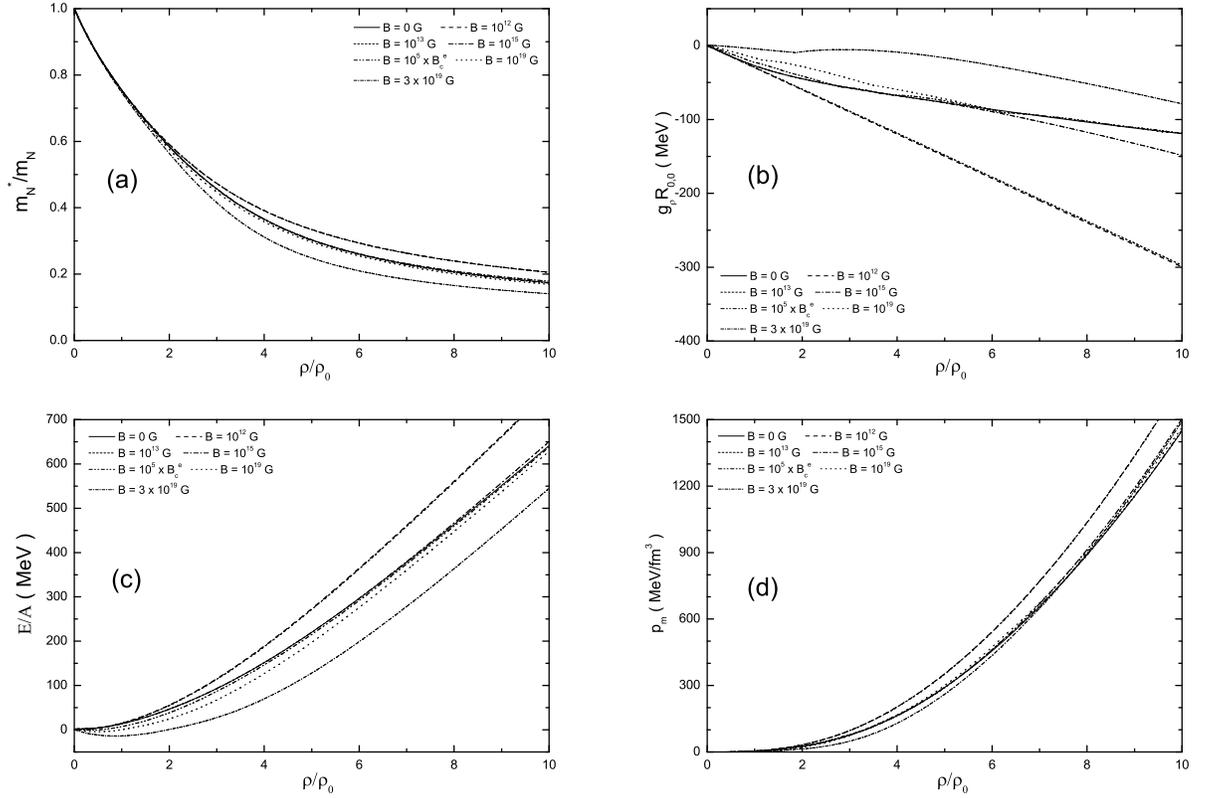,height=0.6\textheight, clip=}}
\caption{The results for n-p-e-$\mu$ system calculated with
SetA($NL\sigma\omega\rho$) without $\delta$-field. (a) shows the
nucleon effective mass as a function of the baryon density for
magnetic fields as presented in Figure2; (b) displays the
$\rho$-field strength $g_{\rho}R_{0,0}$ ; (c) and (d) show the
energy per nucleon and pressure density, respectively. }
\end{figure}
\newpage

\newpage
\begin{figure}[h]
\label {fig6} \centering
\mbox{\epsfig{figure=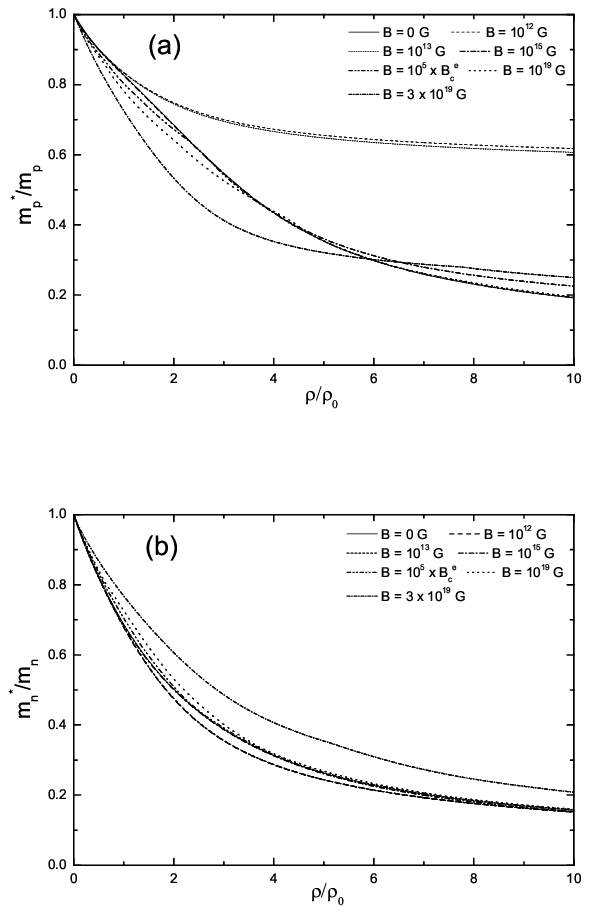,height=0.8\textheight, clip=}}
\caption{Same as Figure 2, except that the AMM terms of nucleons
and muons are included.}
\end{figure}
\newpage
\begin{figure}[h]
\label {fig7} \centering
\mbox{\epsfig{figure=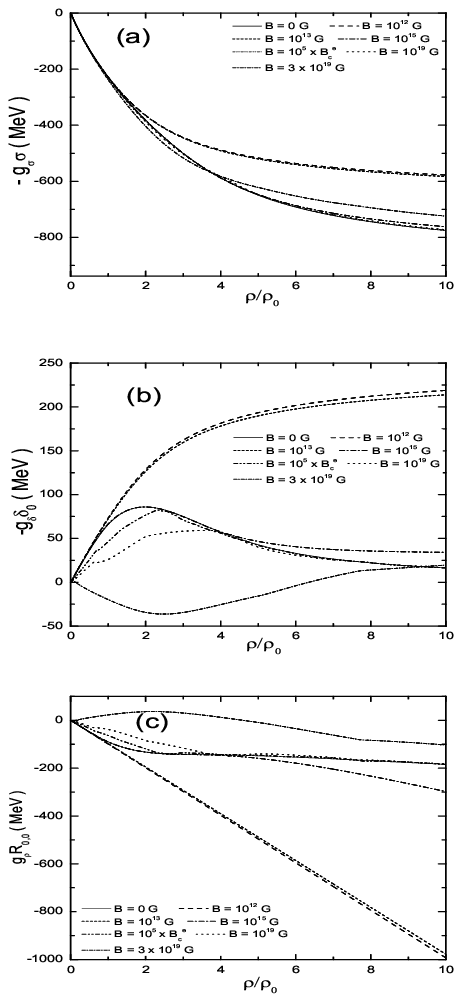,height=0.9\textheight, clip=}}
\caption{Same as Figure 3, except that the AMM terms of nucleons
and muons are included.}
\end{figure}

\newpage
\begin{figure}[h]
\label {fig8} \centering
\mbox{\epsfig{figure=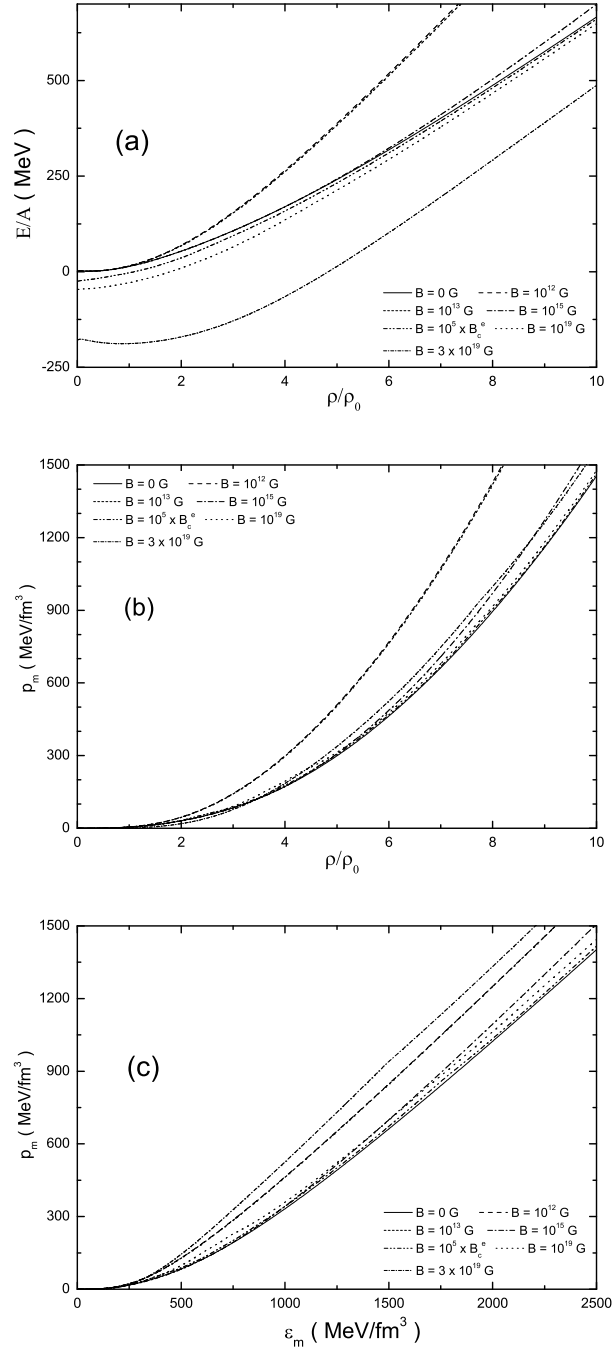,height=0.9\textheight, clip=}}
\caption{ Same as Figure 4, except that the AMM of nucleons and
muons are included.}
\end{figure}
\newpage

\begin{figure}[h]
\label {fig9} \centering
\mbox{\epsfig{figure=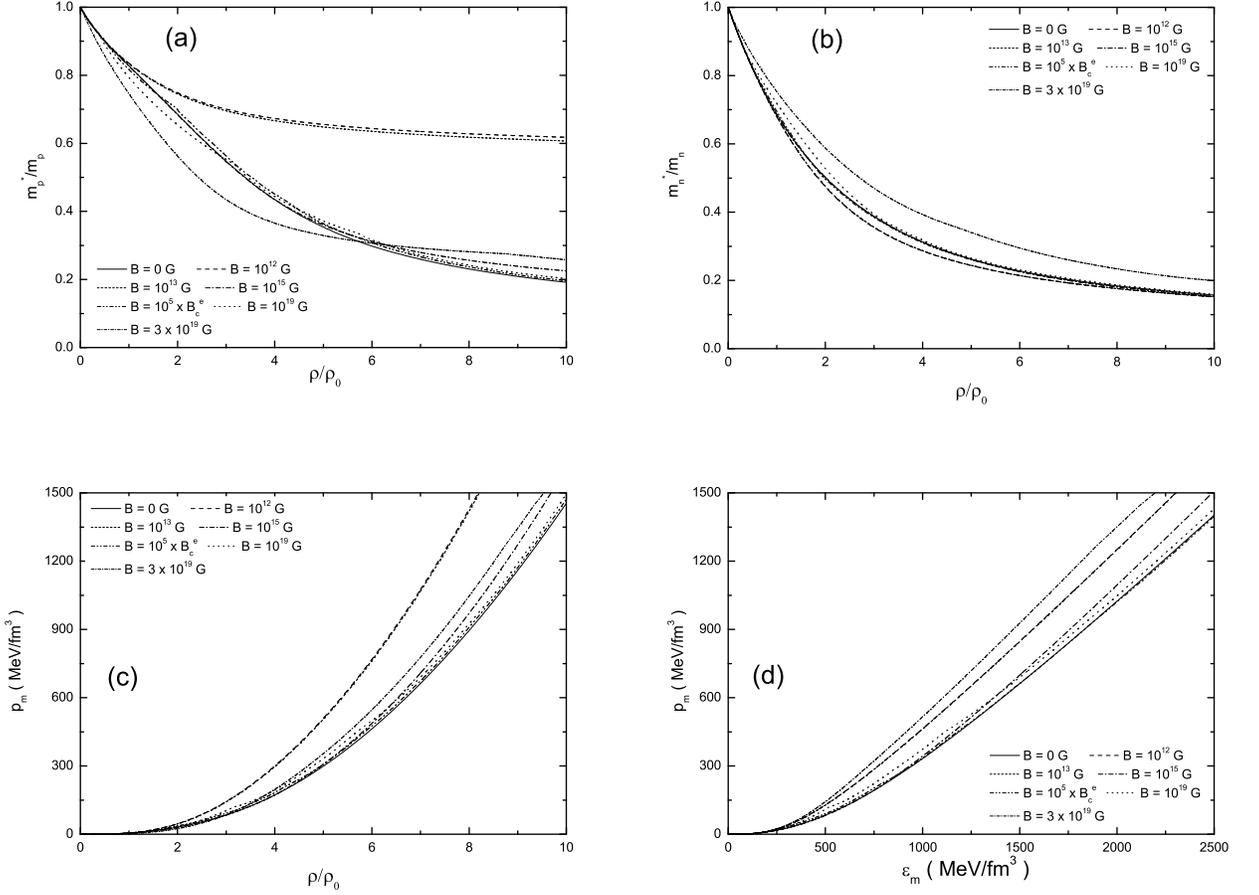,height=0.65\textheight, clip=}}
\caption{Results including the effect of AMM of electron for model
SetA(NL$\sigma\omega\rho\delta$). (a) and (b) present the
effective masses of proton and neutron , respectively; (c) and (d)
show the pressure densities as functions of baryon densities and
energy densities, respectively.}
\end{figure}

\end{document}